# Ultrafast synthesis of SiC nanowire webs by floating catalysts rationalised through in-situ measurements and thermodynamic calculations

Isabel Gómez-Palos,[†,a,b] Miguel Vazquez-Pufleau,[†,a] Jorge Valilla,[a,c] Álvaro Ridruejo,[b] Damien Tourret,[a,*] and Juan J. Vilatela[a,*]

[a] IMDEA Materials, Madrid, 28906, Spain. [b] Department of Materials Science, Universidad Politécnica de Madrid, E.T.S. de Ingenieros de Caminos, 28040 Madrid, Spain. [c] Universidad Carlos III de Madrid, 28911 Leganes, Spain.

*damien.tourret@imdea.org, juanjose.vilatela@imdea.org.

This work presents the synthesis of SiC nanowires floating in a gas stream through the vapour-liquid-solid (VLS) mechanism using an aerosol of catalyst nanoparticles. These conditions lead to ultrafast growth at 8.5 µm/s (maximum of 50 µm/s), which is up to 3 orders of magnitude above conventional substrate-based chemical vapour deposition. The high aspect ratio of the nanowires (up to 2200) favours their entanglement and the formation of freestanding network materials consisting entirely of SiCNWs. The floating catalyst chemical vapour deposition growth process is rationalised through in-situ sampling of reaction products and catalyst aerosol from the gas phase, and thermodynamic calculations of the bulk ternary Si-C-Fe phase diagram. The phase diagram suggests a description of the mechanistic path for the selective growth of SiCNWs, consistent with the observation that no other types of nanowires (Si or C) are grown by the catalyst. SiCNW growth occurs at 1130 °C, close to the calculated eutectic. According to the calculated phase diagram, upon addition of Si and C, the Fe-rich liquid segregates a carbon shell, and later enrichment of the liquid in Si leads to the formation of SiC. The exceptionally fast growth rate relative to substrate-based processes is attributed to the increased availability of precursors for incorporation into the catalyst due to the high collision rate inherent to this new synthesis mode.

## Introduction

There is a strong interest in methods that enable large-scale assembly of high-aspect ratio nanomaterials as macroscopic structures. Macroscopic networks of nanowires, nanotubes and 2D nanomaterials show promise in a myriad of applications, ranging from purification[1] to energy storage[2,3] and conversion[4], from thermal management[5,6] to electronics[7].

There are various routes to synthesise inorganic nanowires in powder form[8,9]. But, to produce macroscopic structures, the prevalent method has been so far through their growth as vertically aligned arrays on a substrate of interest by chemical vapour deposition (CVD)[10,11]. In the process, nanowires grow through the so-called vapour-liquid-solid (VLS) mechanism, whereby gas precursors decompose and get incorporated into a metal catalyst nanoparticle on a substrate forming an alloy, which then, upon supersaturation, segregates a solid nanowire. Nanowire diameter, position and number density are controlled and fixed through pre-deposition of the catalyst nanoparticles[12,13].

Recently, we proposed an alternative route to produce macroscopic structures of nanowires through the synthesis of silicon nanowires floating in a gas stream from an aerosol of gold catalyst nanoparticles in a flow-through reaction. The floating catalyst chemical vapour deposition (FCCVD) method enabled continuous assembly of macroscopic sheets of Si nanowires[14]. A similar method is well known for spinning of continuous fibres of carbon nanotubes[15,16]. Nanowires of III-V semiconductors are also known to grow through this method[17,18,19].

The fascinating prospect is that FCCVD is a new general route for assembly of inorganic nanowires as macroscopic structures that eliminates all processing solvents and substrates, and which leads to a new class of textile-like network materials. This paper sets out to support this hypothesis by demonstrating that silicon carbide nanowires (SiCNW) also can be synthesised by FCCVD and directly form macroscopic freestanding solids.

Silicon carbide nanowires combine the properties of a refractory ceramic with a wide band gap (2.3 eV), high field strength, and high charge mobility. They are thus particularly attractive for high frequency[20], high power[21] and high temperature electronic devices[22], but have also been used for photocatalysis[23], field emission[24], electromagnetic radiation absorption[25], and energy storage[26].

Numerous methods for SiCNW synthesis exist, including template assisted growth[27], laser ablation[28] and carbothermal reduction,[29] but chemical vapour deposition (CVD) is preferred because of its superior yield, selectivity and ease of implementation.[30,31,32] SiCNW synthesis by CVD is generally carried out using transition metal nanoparticles pre-deposited on a silicon wafer as a catalyst. Variations include the use of other metals[33], generation of Fe catalyst in the gas phase[34] and growth on different substrates (e.g. carbon fibre[24]). There are also several forms of introducing precursors of CVD, either as
separate C and Si compounds[35] or as molecules containing both elements, such as polysilacarbosilane[36] and hexametyldisilane[37,38]. Despite this wide range of synthesis conditions, reported SiCNW growth rates by CVD are narrow, and slow. Average growth rate for conventional substrate processes is ≈5 nm/s. Moreover, the presence of a substrate restricts sample format, thus limiting the application of SiCNWs. This work demonstrates that eliminating the substrate from synthesis of SiCNWs increases growth rate by several orders of magnitude and enables the direct assembly of freestanding networks of nanowires resembling fabrics. The paper provides a description of this new synthesis mode based on analysis of reaction products sampled in-situ from the gas phase, and on thermodynamic calculations of phase equilibria for the Fe-Si-C system.

## Experimental

SiC nanowires were synthetized in a vertical flow-through FCCVD reactor via the VLS mechanism, using an aerosol of Fe as the catalyst and hexamethyldisilane (HMDS, $Si_2C_6H_{18}$) as SiC precursor. The window of reaction conditions explored for this study is summarised in Table 1. It also includes the conditions judged optimal in terms of observed conversion, and which are therefore those used in the experiments in this report, unless otherwise noted.

The liquid HMDS is fed through an injector placed at the upper part of the reactor (13 cm below the top flange at 485 °C) at a rate of 0.5 ml/h. The injection system consists of a capillary tube (1/16" OD and 0.010" ID) with a surrounding $H_2$ flow of 60 ml/min to aid the atomization of the liquid. Iron catalyst can be introduced in the reactor by two different means. As a standard case, ferrocene is dissolved in liquid HMDS until saturation (≈16 mg/ml). Alternatively, Fe nanoparticles are formed with a high voltage spark generator (SG) using iron electrodes (98.5 % purity) and carried to the reactor with a $H_2$ flow[39]. The reactor tube (mullite 660, $Ø_{int}$ = 7 cm) was set at a nominal temperature of 1300 °C and a $H_2$ atmosphere is sustained by a 3 lpm $H_2$ flow.

Along the reactor tube, and particularly near the hottest zone of the reactor, visible solid products are formed.

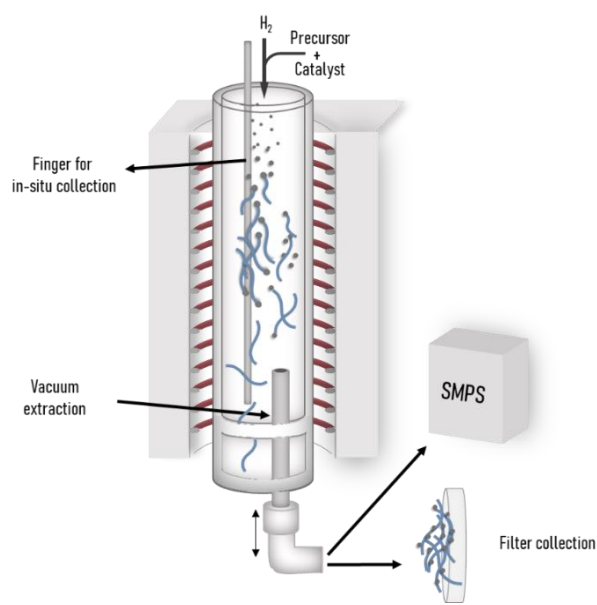

**Table 1**  Explored experimental variables in the synthesis of SiCNWs.

| Variable | Range explored | Optimal |
|---|---|---|
| Nominal temperature | 1200 – 1300 °C | 1300 °C |
| Carrier gas flow | 0.32 – 3 lpm | 3 lpm |
| % H2 in carrier gas | 0 – 100 % | 100 % |
| Precursor injection rate | 0.5 – 2 ml/h | 0.5 ml/h |
| Ferrocene concentration | 1 – 16 mg/ml | 16 mg/ml |

**Fig. 1**  Schematic of the reactor for the synthesis of SiCNWs through floating catalyst CVD. It includes a ceramic finger for in-situ collection of nanomaterials, and a system for vacuum extraction of floating nanowires or catalyst aerosol nanoparticles.

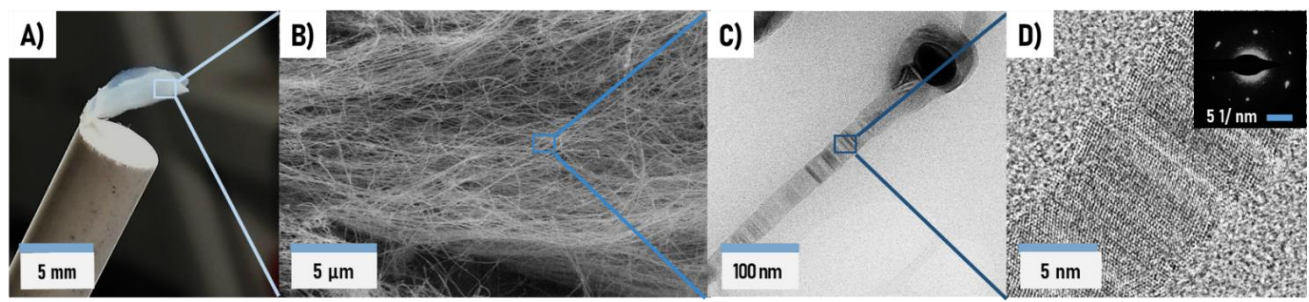

**Fig. 2** SiCNW materials produced by FCCVD and collected as solids. A) Photograph of an aerogel of entangled SiCNWs forming a freestanding solid. B) SEM micrograph showing the network structure of high-aspect ratio SiCNWs. C) TEM micrograph of a SiCNW with the typical VLS growth tip and stacking faults. D) HRTEM micrograph showing the nanowire 3C crystal structure. The FFT is shown in the insert. (SiCNWs synthesised at standard conditions and 20 min of collection)

Two main systems were used to collect solid products from the gas phase. The first one consists of a thin mullite rod (6 mm diameter) inserted through the whole length of the reactor and kept there during the reaction. It has no form of cooling and can thus be assume to have the same temperature profile as the reactor tube. The second consists of a vacuum extraction system with controllable position to sample material from different parts of the reactor chamber. The system consisted in a 1-m-long ceramic tube (1/4" diameter) extracting 0.5 lpm of the gas inside the reactor by means of a Venturi pump-critical orifice[40] and collected with a polyethersulfone (PES) filter at the end. This movable extraction system was also used to determine the evolution of the Fe aerosol. It was measured in real time with a scanning mobility particle sizer (SMPS, GRIMM MODEL 5705) which selects particle size based on a differential mobility analyser (DMA, GRIMM MODEL 5706) to obtain data of the concentration and size distribution of the catalyst. **Fig. 1** shows a schematic representation of the experimental set-up and collection systems. The ceramic rod and vacuum filter provide separate means to sample and then analyse the morphology and composition of solid products, however we note that these collection systems are probably size selective (See SI).

After extraction, samples were characterized by optical imaging and scanning electron microscopy (SEM, FEI Helios NanoLab 600i). Micrographs were used to obtain the diameter and length distribution by image analysis (see SI). The distributions of diameter and length were obtained from inspection of 500 and 170 NWs, respectively. Raman spectroscopy (Raman Spectrometer RENISHAW MoD. Invia) was carried out with a laser wavelength of 532 nm (2.33 eV), using 50x objectives. High resolution transmission electron microscopy (HRTEM, FEI Talos F200X) and synchrotron x-ray diffraction at ALBA synchrotron facilities (NCD-SWEET beamline) were performed to assess the crystallinity and phase composition of the nanowires.

In addition, we calculated the ternary Fe-Si-C equilibrium phase diagram at different temperatures using the CalPhaD method, in order to rationalise the compositional path followed from the Fe catalyst to the growth of SiC via addition of HMDS. These calculations were performed with software ThermoCalc® and the thermodynamic database for Fe-based alloys TCFE9.

## Results

**Nanowire characterization**

Samples consisting of SiCNW webs were successfully synthesised via FCCVD without any substrates and directly collected from the growth zone of the reactor using either a ceramic rod or a vacuum extraction system. **Fig 2** presents optical and electron micrographs of standard samples showing the hierarchical structure of these materials. The nanowires are highly crystalline, as shown in the HRTEM image of **Fig 2c-d.** The interplanar spacing of the crystalline SiC is d = 0.25 nm, coinciding with the (111) plane of 3C-SiC cubic structure (JCPDS Card no. 73-1665). Stacking faults (SF) a few nanometres long are observed parallel to the growth axis. These SF, formed from dislocation in the atomic lattice, are commonly observed in nano-SiC and allow the formation of nanosegments of different SiC phases[28].

Raman spectra in **Fig. S1** shows the characteristic peeks at 790 and 940 cm$^{-1}$ corresponding to the transverse optic (TO) and longitudinal optic (LO) mode of SiC. The peak at 940 cm$^{-1}$ is shifted from the bulk peak (972 cm$^{-1}$) due to the nanometre dimensions of the SiCNWs[41]. Other peaks found at 870 and 760 cm$^{-1}$ are associated to structural defects in the materials, in this case related to the stacking faults found in the nanowires[42]. Crystallinity of the SiC NWs over large sample volumes was confirmed with synchrotron X-ray diffraction (see **Fig. S2**). The material is composed predominantly of the 3C phase of SiC, with weaker peaks corresponding to the 6H-SiC phase. This signal confirm that nanosegments of different phases are generated by the stacking faults[41]. Application of the Scherrer equation in the three main peaks of the 3C-SiC phase result in an average crystallite size of 19.6 nm. The tip catalyst composition after cooling was assessed by EDS and HRTEM (Fig. **S3**). The catalyst cores observed consisted of the $Fe_3Si$ intermetallic. The whole catalyst tip was found to be composed of Fe, O and Si, with some traces of C, similar to other report on SiC nanowire growth using Fe as catalyst. [43,44,45,46]

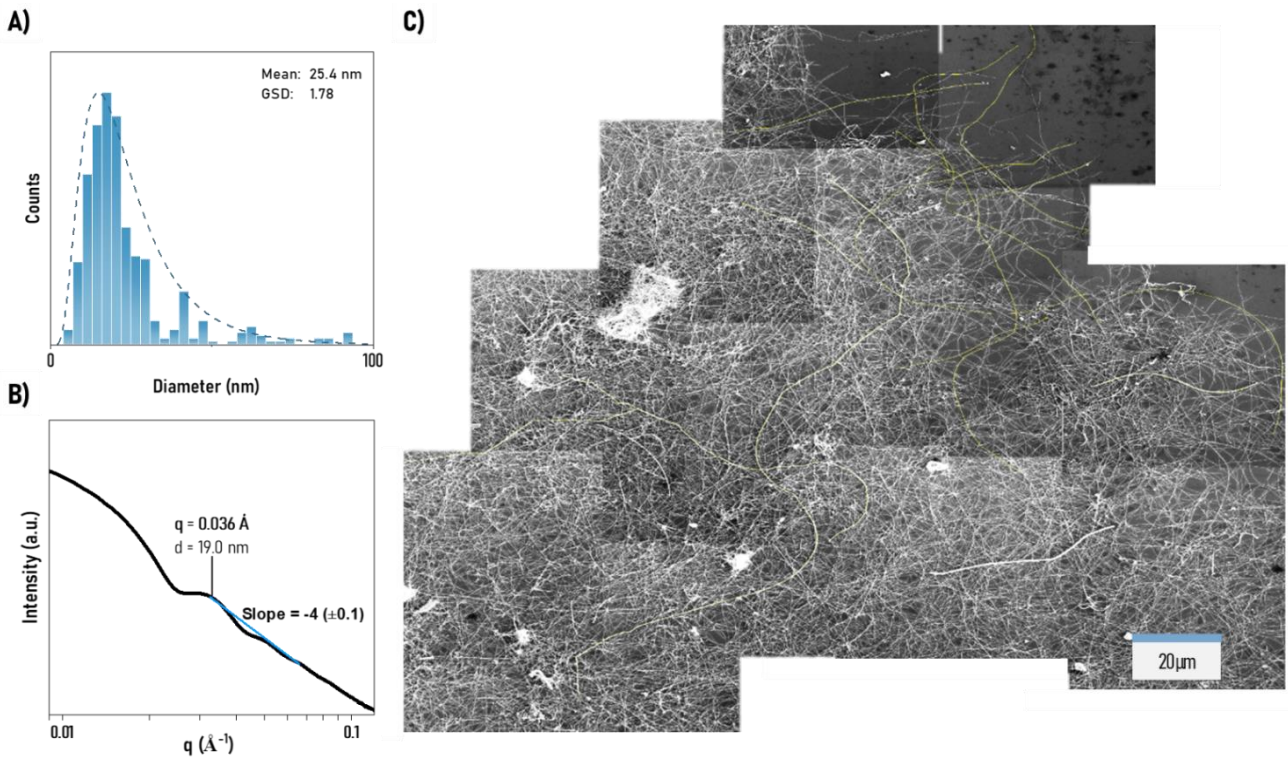

**Fig. 3** Structure of SiCNW networks produced by FCCVD. A) Diameter distribution measured by SEM images analysis of different samples synthesised at standard conditions. B) SAXS intensity profile showing low polydispersity in a single self-standing sample. C) SEM micrograph of a sample collected from the gas-phase for determination of nanowire length.

The macroscopic solid produced by this method is a network material, whose bulk properties are often not dominated by intra-particle properties, but rather through their spatial organisation and shape. Hence, there is strong interest in determining the morphology of the nanowires. The formation of a freestanding solid is possible due to the exceptionally high aspect ratio of the SiCNWs (up to 2200 with an average of 773), which favours their entanglement in the gas phase, as also observed for other inorganic nanomaterials[14,47]. For reference, the high aspect ratio of nanowires grown by FCCVD reported are 214 for SiNW[14] and 12.5 for GaAs[17]. From SEM image analysis of sparse networks, we found NW lengths around 20 µm with a maximum value of 123 µm (**Fig. 3c,** for more details in length measurement see SI). Diameter size distribution was obtained from image analysis of SEM micrographs of 500 nanowires , including from different samples. **Fig. 3a** shows the experimental data, which has a log-normal distribution with diameters ranging from 2 to 100 nm, a mean value of 25.4 nm and geometric standard deviation σ = 1.78. The nanowire diameter distribution stems from the size distribution of the catalyst, which is also lognormal. Moreover, metal particle aerosols are intrinsically more polydisperse (σ > 1.36) than metal particles produced on a substrate (σ < 1.34).[48] These results confirm that the SiCNWs grow from floating catalyst particles, and suggest that their diameter distribution could be made narrower through catalyst selection methods, albeit at the expense of a lower particle concentration.

Interestingly, the SiCNW network material gives rise to strong features in SAXS. Figure **3b** shows the radial intensity profile for a self-standing sample. The intensity profile shows clear peaks at mid values of scattering vector ($q$), with a particularly intense one at a d-spacing of 19 nm. We attribute these features to the form factor of the SiCNWs. Their prominence over the Porod slope suggest smaller diameter polydispersity than in other nanowire ensembles studied previously.[49]

These results were obtained using ferrocene as catalyst source and in a $H_2$ atmosphere. Interestingly, a mixture of $H_2$ and $N_2$ produced higher concentration of nanoparticles without detectable Fe catalyst content. An example is included in **Fig. S4**.

**Gas-phase sampling of reaction products**

The interest then is in determining and rationalising the synthesis conditions leading to selective, fast growth of SiC nanowires by FCCVD. The experimental conditions used in this growth mode correspond to a flow-through reactor, characterised by two main features. Firstly, that both the catalyst and SiC precursor evolve as they travel through the hot tubular furnace. Second, that residence time is very short, of less than 20 s.

To gain more insight into the growth mechanism, we equipped the flow-through reactor with two probes for collection of solid reaction products directly from the reaction zone. The premise is that the material collected at different positions along the reactor tube reflects the progression of the reaction through it.

Using a ceramic rod introduced in the reaction zone, we could directly collect solid products and relate them to the reaction temperature. In a separate experiment, we could also determine the evolution of the size and concentration of the catalyst aerosol produced with the SG (in the absence of SiC precursors). The correspondence between solid products, temperature profile and catalyst evolution is shown in **Fig. 4**.

This analysis indicates that growth of SiCNWs occurs only in a small region of ≈3 cm in the hot zone, where the temperature reaches 1120 - 1140 °C. The SiCNWs grown at this position form a clear bluish aerogel around the ceramic rod (Fig. 4). Similarly, using a vacuum extraction system, SiCNWs were only found when the extraction probe was positioned below this reaction zone, i.e., deeper into the reactor in the direction of the flow (**Fig. S5**). Furthermore, the same growth zone was observed independently of the source of Fe catalyst. SiCNW growth concentrated in the range 1120 - 1140 °C using Fe catalyst nanoparticles produced either with a spark discharge generator or from the decomposition of ferrocene (**Fig. S6**).

The concentration and size distribution of the catalyst generated with the high voltage spark instrument was assessed with an SMPS along the reactor. The colour map depicted in **Fig. 4** shows that the mean particle number and size remain fairly constant along the whole reactor path, at $1.5 \times 10^7$ particles/cm$^3$ and 10 nm, respectively. Essentially, the catalyst particles produced with the SG have a suitable size for SiCNW growth all the way from their entry into the reaction tube until the exit at the outlet and the diameter distribution of the nanowires produced coincides with the catalyst size (Fig. **S7 A**). The average particle concentration and size distribution of the nanoparticles generated by decomposition of ferrocene are calculated to be $4.2 \times 10^8$ particles/cm$^3$ using the size distribution of the nanowires produced with this source. Fig. **S7 B** shows a comparison in the diameter distribution of the nanowires using both catalyst sources.

In addition to SiCNWs, the sampling probes used show other nanostructured solid products of the reaction. However, their growth is not mediated by the Fe catalyst. No traces of Fe were found in these solid products. When using the ceramic rod, microscopic aggregates of Si were found strongly attached to the ceramic in a region corresponding to a temperature of approximately 1050°C. They were only detected in the presence of the ceramic tube and contain no Fe, thus, most likely they form through precursor decomposition[50] and solid nucleation on the porous mullite surface. Similarly, soot particles are produced and without mediation of the catalyst. They are found with both sampling probes, implying that they form through pyrolysis in the gas phase (for more detailed information about the by-products, see SI).

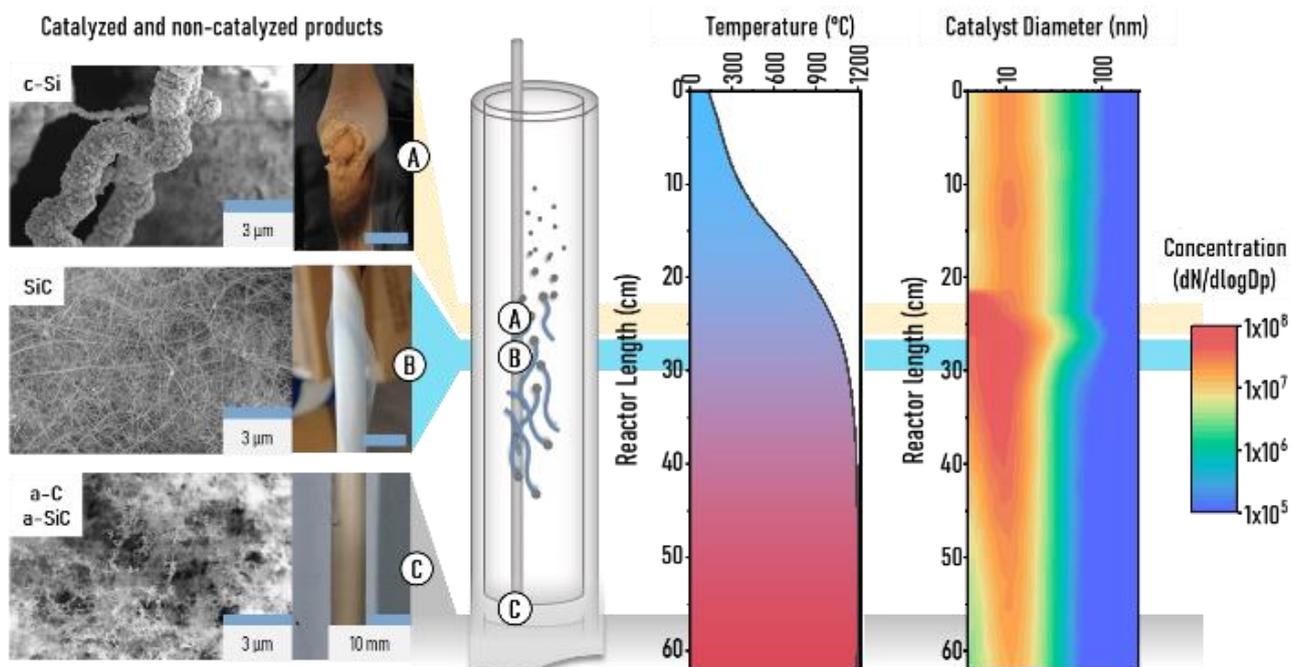

**Fig. 4** Reactions in the synthesis of SiCNW by FCCVD. Collected materials synthesised through the flow reactor, the corresponding temperature profile and catalyst aerosol evolution in the reaction area.

A set of control experiments were also conducted to confirm that the sampled NWs grow through floating catalyst, rather than on the ceramic sampling rod, for example. In the first experiment, catalyst particles were produced in the absence of SiC precursors and their deposition on the ceramic rod was analysed. It was found that they attached to the ceramic rod only in sections at temperatures below 600 °C and far from the region where NWs are observed under FCCVD (Fig. **S8**). In another experiment, NWs were synthesised without inserting any sampling parts in the reaction zone, and collected at the outlet with a vacuum system. The SiCNWs collected with this method were identical to those sampled directly in the reaction zone (Fig. **S5E**), although their concentration was lower due to losses to the walls and the presence of soot particles. Finally, we collected NWs with a ceramic rod inserted in the reaction zone for different lengths of time (Fig. **S10**). These last experiments showed almost identical length distributions for different dwell times, and confirmed growth of > 5 µm-long NWs for sampling times as low as 10 s.

## Discussion

A direct outcome of these experiments is a well-defined growth rate. Nanowire growth is only observed over a region of 3 cm, corresponding to a residence time of ≈2.3 s. Taking the nanowire length distribution, the average growth rate comes out as 8.5 µm/s and as high as 50 µm/s. For reference, this is more than three orders of magnitude faster than conventional substrate growth of SiC, (**Fig. 5, Table S1**). It is also far above the highest rates observed using growth enhancing agents such as rare earths (1 nm/s )[51] and Ga (185 nm/s)[52].

These results are in line with previous studies on growth of SiNW[14] and GaAsNW[17], and confirm that nanowire growth by FCCVD is inherently orders of magnitude faster than in substrate-based CVD. This is attributed to faster precursor transport because in FCCVD growth the catalyst is completely surrounded by an excess of gas precursors, increasing the availability for precursor incorporation due to a higher collision rate between catalyst and intermediate active species.

The question then moves onto the kinetic and thermodynamic factors in this synthetic route that control selectivity towards SiCNW growth.

Studies on methyl-substituted silanes clearly show molecular hydrogen and methyl radical as among the first products of the thermal decomposition[53]. Reports on the pyrolytic decomposition of HDMS propose complex decomposition routes, with a large number of decomposition products and intermediates, and multiple pathways[54,55,56]. There is consensus about some initial routes for the decomposition of HMDS:

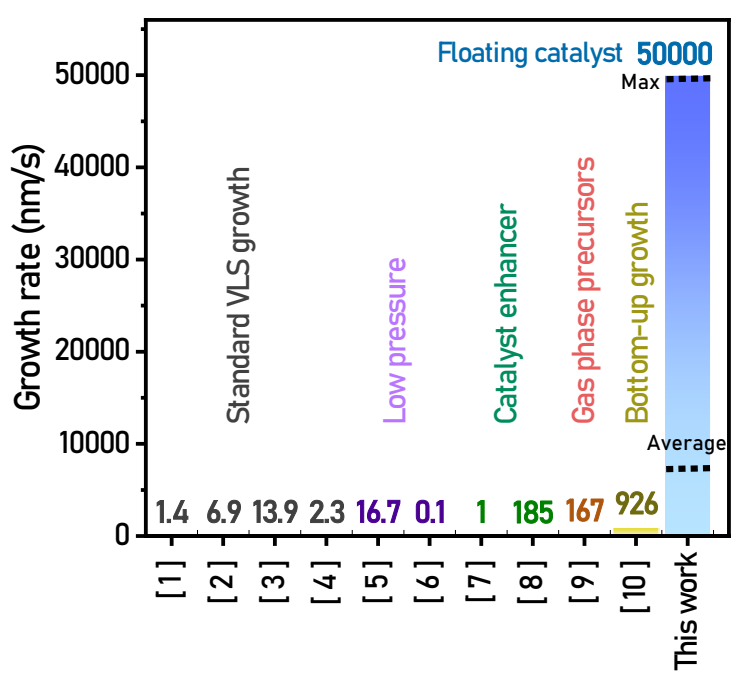

**Fig. 5** Comparison of SiCNW growth rates obtained experimentally. Values were calculated dividing the average length of the nanowires by the experimental growth time. [1] Meng *et al.*[65] [2] Park *et al.*[66] [3] Panda *et al.*[37] [4] Wu *et al.*[24] [5] Zhang *et al.*[67] [6] Zhou *et al.*[68] [7] Rajesh *et al.*[51] [8] Kim *et al.*[52] [9] Attolini *et al.*[30] [10] Li *et al.*[36]

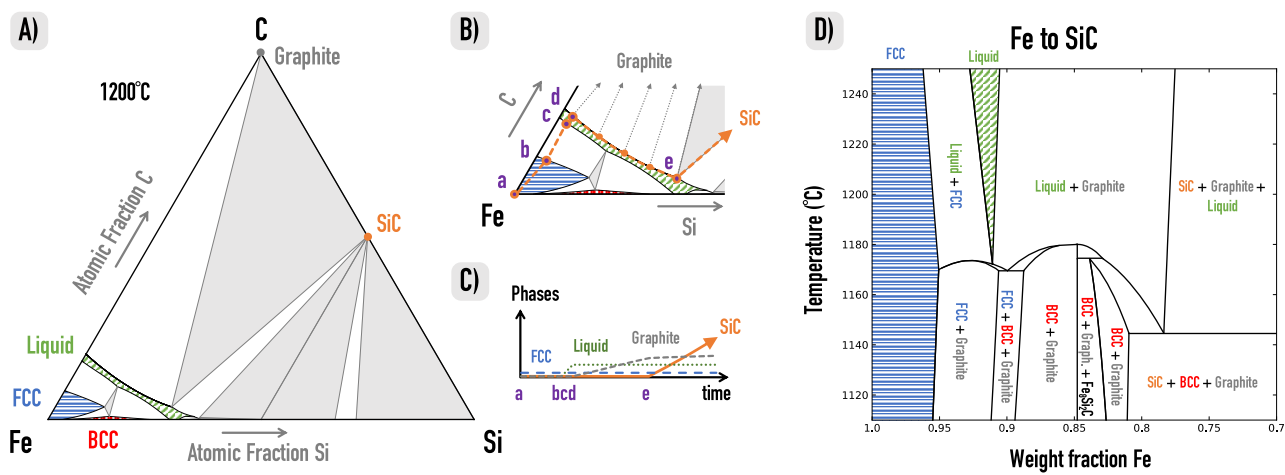

**Fig. 6** Calculated ternary Fe-Si-C phase diagram and compositional path from Fe catalyst to SiC. A) Ternary isothermal phase diagram at 1200°C. Gray triangles show three-phase invariant equilibria; hatched regions show single-phase regions; for clarity, tie-lines are not shown (hence two-phase regions appear white). B) Compositional path from pure Fe (a) to SiC formation (e) following continuous addition of Si and C. C) Qualitative formation of phases along the proposed compositional path. D) Isopleth section of the phase diagram through pure Fe and SiC.

$$Me_3Si - SiMe_3 \rightarrow 2Me_3Si \quad (1)$$
$$Me_3Si \rightarrow Me_2Si + Me \quad (2)$$
$$Me_2Si \rightarrow C_2H_m + SiH_2 \quad (3)$$

Where Me is a methyl group (-CH$_3$) and *m* is the number of H atoms, which can be 2 or 4. Note that for m=2, H$_2$ would be produced in (3).

The methyl and silylene radicals are precursors in nanoparticle formation (a competitive route to nanowire formation) and form them in a simplified reaction mechanism as follows:

$$CH_4 \rightarrow Me + C_xH_y \rightarrow C_{Nanoparticle} \quad (4)$$
$$SiH_4 \rightarrow SiH_2 + Si_pH_q \rightarrow Si_{Nanoparticle} \quad (5)$$

In this way, methyl and silylene radicals polymerize by insertion in its parent precursor until forming nanoparticles. The exact reaction mechanism for the formation of C and Si nanoparticles is still under debate[57] and is beyond the scope of this paper. Eq (4) and (5) only show that the path towards C and Si nanoparticles involve the formation of radicals of SiH$_4$ and CH$_4$. The methyl and silylene radicals can also react with hydrogen in the following way:

$$Me + \tfrac{1}{2} H_2 \rightarrow CH_4 \quad (6)$$
$$SiH_2 + H_2 \rightarrow SiH_4 \quad (7)$$

Considering the reaction paths (6) and (7), a hydrogen atmosphere inhibits nanoparticle formation, slowing down their nucleation by neutralising methyl and sylene radicals. This allows the precursors to have a larger lifespan enhancing their chance of impinging into Fe catalysts to form SiC nanowires. A hydrogen atmosphere is thus expected to reduce the effective rate of decomposition of silylene and methyl radicals, following Le Chatelier's principle, and thus prevent the non-catalysed formation of Si, C and SiC nanoparticles. This role of hydrogen is similar to that observed in FCCVD growth of CNTs[47] and SiNWs[14,58]. In the latter, hydrogen avoids non-catalysed nucleation of amorphous nanoparticles by preventing polymerisation of silane derivates[59].

Similarly, we can rationalise the formation of different solid products, both catalysed and non-catalysed, by considering the temperature profile in this FCCVD process (Fig4). HMDS is expected to dissociate at a fast rate compared with typical reactor residence times, at temperatures of 700 °C[56]. At the low Si/H$_2$ molar concentration used in this study (≈0.03 % HMDS), Si nanoparticles formation from silane in H$_2$ would start at around 850°C.[60] For comparison, methane decomposition in H$_2$ to form C soot requires temperatures above 1000 °C.[61] C$_2$H$_4$, C$_2$H$_2$ and other secondary products of HMDS may form C soot at lower temperatures, but generally above those found for Si nanoparticle formation.

In summary, seen in light of the decomposition of HMDS, selective growth of SiCNW requires the inhibition of nanoparticle formation, which is a side reaction that depletes the reagents for NW formation. This is achieved by utilizing hydrogen atmosphere, which reduces the reaction rate of the side reaction and allows catalysed growth before the onset of nanoparticle nucleation, which would otherwise occur first for Si and then for C.

It is also important to consider that in the thermal decomposition of HDMS all initial products and intermediates, and most secondary products, are C-containing silanes and silylenes. This suggests that the species that impinge on the catalyst could contain both Si and C for the subsequent SiC extrusion.

Under the premise that the Fe catalyst receives a continuous influx of Si and C, we can produce a qualitative description of the nucleation of the SiCNW based on the thermodynamics of the Si-C-Fe system.

**Fig. 6** presents the CalPhaD-calculated Fe-Si-C equilibrium ternary phase diagram at 1200 °C, together with a proposed reaction pathway within the diagram, and an isopleth Fe-SiC section through pure Fe and SiC compound. We note that these calculations show the equilibrium phases for a bulk system, neglecting interfacial energy contributions that are not usually negligible in nanosized systems. When accounted for, these contributions tend to shift the phase diagram toward lower temperatures[62,63,64]. This may explain why the temperature window assessed experimentally, over a narrow length of 3 cm at a nearly constant temperature of 1130 ˚C, is lower than the calculated ≈1170˚C (**Fig. 6d**). However, even if calculations remain qualitative, they allow the identification of expected phases and a rationalisation of the reaction pathway.

The dashed line in **Fig. 6b** illustrates the compositional path starting from a Fe-rich particle (labelled a) exposed to an influx of Si and C, assuming a C:Si influx ratio of 1:3 from HMDS. First, both Si and C enter in solution within the FCC Fe phase. At their limit of solubility (b), a liquid phase forms (c), which will also get enriched in C and Si until the limit of solubility within the liquid is reached (d). Then, the excess C in the liquid leads to the formation of graphite, and upon further addition of Si and C, the liquid composition follows the Liquid-Graphite solvus line until reaching the point of nucleation of SiC compound (e).

This analysis has several interesting corollaries of interest to explain the chemical composition of products of the reaction.

The ternary diagram shows that SiC will form irrespectively of the initial compositions of the Si-C species impinging on the Fe catalyst since the liquid phase exhibits only equilibria with graphite and SiC phases along its Fe-lean solvus line at T ≈ 1200˚C. The initial ratio of Si to C in the liquid will simply shift the fraction of SiC to $C_{(solid)}$ in the final product. The resulting reaction path (phase fraction vs time) is schematically represented in **Fig. 6c**. The compositions of points (a) through (e) are listed in **Table S2**. Notably, the C:Si ratio of point (e) is 2.54:1, which is relatively close to that of HDMS (3:1), suggesting that the reaction and growth of SiC can proceed from point (e) with minimal shift of the liquid composition under the influx of Si and C via HMDS.

In addition, we note that this phase evolution appears to be the same for all temperatures above the eutectic, as shown in the Fe-SiC isopleth (**Fig. 6d**). This indicates that the thermodynamic product of a Fe catalyst nanoparticle supersaturated with C and Si, is a SiC nanowire.

Finally, it seems important to note that, in spite of neglecting nano-size effects (surface energies) and using a thermodynamic database originally developed for steel metallurgy, we were able to identify the SiCNWs reaction pathway. This opens a promising avenue for the exploration of a broader range of databases for nanomaterial synthesis. A proper integration of surface effects may even allow a more quantitative approach for the design and synthesis of new NW.

## Conclusions

This work shows a method to synthesise SiCNWs through floating catalyst chemical vapour deposition in a flow-through reactor system. This growth mode leads to an exceptionally fast rate, 2 - 3 orders of magnitude above substrate-based processes, due to the high collision rate of precursors and the catalyst nanoparticles aerosol. The NWs reach very high aspect ratios (1800) and entangle to form freestanding solids solely composed of SiCNWs.

The growth of SiCNWs is rationalised through analysis of reaction products as they evolve through the reactor. SiCNWs are the only solid produced from the Fe catalyst. They were grown in a hydrogen atmosphere which inhibits the competitive formation of side products (Si, C and SiC nanoparticles) during the pyrolysis of the SiC precursor, allowing the incorporation of Si-C species into the catalyst.

SiCNW growth takes place at a temperature 1130 °C, which is close to the expected eutectic point after consideration of size effects. The calculated ternary equilibrium Fe-Si-C phase diagram was used to rationalise the selective growth of SiCNWs. The diagram shows that SiC is the expected thermodynamic product of a saturated liquid of Fe-Si-C, for all practical compositions.

These results give further evidence to support the view that virtually all 1D inorganic nanomaterials can be synthesised by floating catalyst chemical vapour deposition and directly assembled as macroscopic network solids.

## Conflicts of interest

There are no conflicts to declare

## Acknowledgements

The acknowledgements come at the end of an article after the conclusions and before the notes and references. The authors thank the European Union Horizon 2020 Programme under grant agreement 101045394 (ERC-2021-COG, UNIYARNS), by the Madrid Regional Government (FOTOART-CMP2018/NMT-4367 and UPM/APOYO-JOVENES- F6TCCN-145-GGT34M) and Marie Sklodoska Curie Fellowship SUPERYARN under grant number: 101029091 for generous financial support. The authors thank NCD-SWEET beamline staff at ALBA Synchrotron Light Facility for assistance with synchrotron experiments. D.T. acknowledges support by the Spanish Ministry of Science through a Ramón y Cajal Fellowship (RYC2019-028233-I).

## References

1  D. T. Schoen, A. P. Schoen, L. Hu, H. S. Kim, S. C. Heilshorn and Y. Cui, *Nano Lett.*, 2010, **10**, 3628–3632.
2  L. Hu and Y. Cui, *Energy Environ. Sci.*, 2012, **5**, 6423–6435.
3  J. W. Choi, L. Hu, L. Cui, J. R. McDonough and Y. Cui, *J. Power Sources*, 2010, **195**, 8311–8316.
4  W. Dong, A. Cogbill, T. Zhang, S. Ghosh and Z. R. Tian, *J. Phys. Chem. B*, 2006, **110**, 16819–16822.
5  J. Chen, X. Liao, M. Wang, Z. Liu, J. Zhang, L. Ding, L. Gao and Y. Li, *Nanoscale*, 2015, **7**, 6374–6379.
6  D. Lu, L. Su, H. Wang, M. Niu, L. Xu, M. Ma, H. Gao, Z. Cai and X. Fan, *ACS Appl. Mater. Interfaces*, 2019, **11**, 45338–45344.
7  X. Zhang, W. Lu, G. Zhou and Q. Li, *Adv. Mater.*, 2020, **32**, 1902028.
8  Y. Xia, P. Yang, Y. Sun, Y. Wu, B. Mayers, B. Gates, Y. Yin, F. Kim and H. Yan, *Adv. Mater.*, 2003, **15**, 353–389.
9  C. Buzea and I. Pacheco, in *Handbook of Nanofibers*, Springer International Publishing, Cham, 2019, pp. 557–618.
10 L. Güniat, P. Caroff and A. Fontcuberta I Morral, *Chem. Rev.*, 2019, **119**, 8958–8971.
11 Y. J. Hong and C.-H. Lee, *Semiconductor Nanowires I - Growth and Theory*, 2015, vol. 93.
12 S. A. Dayeh and S. T. Picraux, *Nano Lett.*, 2010, **10**, 4032–4039.
13 A. Klamchuen, T. Yanagida, M. Kanai, K. Nagashima, K. Oka, S. Rahong, M. Gang, M. Horprathum, M. Suzuki, Y. Hidaka, S. Kai and T. Kawai, *Appl. Phys. Lett.*, 2011, **99**, 58–61.
14 R. S. Schäufele, M. Vazquez-Pufleau and J. J. Vilatela, *Mater. Horizons*, 2020, **7**, 2978–2984.
15 A. Mikhalchan and J. J. Vilatela, *Carbon N. Y.*, 2019, **150**, 191–215.
16 L. Weller, F. R. Smail, J. A. Elliott, A. H. Windle, A. M. Boies and S. Hochgreb, *Carbon N. Y.*, 2019, **146**, 789–812.
17 M. Heurlin, M. H. Magnusson, D. Lindgren, M. Ek, L. R. Wallenberg, K. Deppert and L. Samuelson, *Nature*, 2012, **492**, 90–94.
18 W. Metaferia, A. R. Persson, K. Mergenthaler, F. Yang, W. Zhang, A. Yartsev, R. Wallenberg, M. E. Pistol, K. Deppert, L. Samuelson and M. H. Magnusson, *Nano Lett.*, 2016, **16**, 5701–5707.
19 S. Sivakumar, A. R. Persson, W. Metaferia, M. Heurlin, R. Wallenberg, L. Samuelson, K. Deppert, J. Johansson and M. H. Magnusson, *Nanotechnology*, 2021, **32**, 025605.
20 R. J. Trew, J.-B. Yan and P. M. Mock, *Proc. IEEE*, 1991, **79**, 598–620.
21 Y. Mikamura, K. Hiratsuka, T. Tsuno, H. Michikoshi, S. Tanaka, T. Masuda, K. Wada, T. Horii, J. Genba, T. Hiyoshi and T. Sekiguchi, *IEEE Trans. Electron Devices*, 2015, **62**, 382–389.
22 G. Müller, G. Krötz and E. Niemann, *Sensors Actuators A. Phys.*, 1994, **43**, 259–268.
23 T. Yang, X. Chang, J. Chen, K. C. Chou and X. Hou, *Nanoscale*, 2015, **7**, 8955–8961.
24 R. Wu, K. Zhou, J. Wei, Y. Huang, F. Su, J. Chen and L. Wang, *J. Phys. Chem. C*, 2012, **116**, 12940–12945.
25 S. C. Chiu, H. C. Yu and Y. Y. Li, *J. Phys. Chem. C*, 2010, **114**, 1947–1952.
26 Z. Zhang, C. Fang, J. Muhammad, J. Liang, W. Yang, X. Zhang, Z. Rong, X. Guo, Y. Jung and X. Dong, *Ionics (Kiel).*, 2021, **27**, 2431–2444.
27 G. Xi, Y. He and C. Wang, *Chem. - A Eur. J.*, 2010, **16**, 5184–5190.
28 W. Shi, Y. Zheng, H. Peng, N. Wang, C. S. Lee and S.-T. Lee, *J. Am. Ceram. Soc.*, 2000, **83**, 3228–3230.
29 X. . Li, L. Liu, Y. . Zhang, S. . Shen, S. Ge and L. C. Ling, *Carbon N. Y.*, 2001, **39**, 159–165.
30 G. Attolini, F. Rossi, M. Negri, S. C. Dhanabalan, M. Bosi, F. Boschi, P. Lagonegro, P. Lupo and G. Salviati, *Mater.*

*Lett.*, 2014, **124**, 169–172.
31  Y. Chu, Q. Fu, Z. Zhang, H. Li, K. Li and Q. Lei, *J. Alloys Compd.*, 2010, **508**, L36–L39.
32  X. Qiang, H. Li, Y. Zhang, S. Tian and J. Wei, *Mater. Lett.*, 2013, **107**, 315–317.
33  Q. Pang, L. Xu, Z. Ju, Z. Xing, L. Yang, Q. Hao and Y. Qian, *J. Alloys Compd.*, 2010, **501**, 60–66.
34  Y. Zhang, N. Wang, R. He, X. Chen and J. Zhu, *Solid State Commun.*, 2001, **118**, 595–598.
35  C. Guo, L. Cheng, F. Ye and Q. Zhang, *Materials (Basel).*, 2020, **13**, 1–12.
36  G. Li, X. Li, Z. Chen, J. Wang, H. Wang and R. Che, *J. Phys. Chem. C*, 2009, **113**, 17655–17660.
37  S. K. Panda, J. Sengupta and C. Jacob, *J. Nanosci. Nanotechnol.*, 2010, **10**, 3046–3052.
38  K. F. Cai, Q. Lei and A. X. Zhang, *J. Nanosci. Nanotechnol.*, 2007, **7**, 580–583.
39  M. Vazquez-Pufleau, I. Gomez-Palos, L. Arévalo, J. García-Labanda and J. J. Vilatela, *Nano Res.*, 2022, (Submitted).
40  M. Vazquez-Pufleau, *Powder Technol.*, 2022, **412**, 117974.
41  M. Bechelany, A. Brioude, D. Cornu, G. Ferro and P. Miele, *Adv. Funct. Mater.*, 2007, **17**, 939–943.
42  S.-L. Zhang, B.-F. Zhu, F. Huang, Y. Yan, E. Shang, S. Fan and W. Han, *Solid State Commun.*, 1999, **111**, 647–651.
43  F. J. Narciso-Romero and F. Rodríguez-Reinoso, *J. Mater. Sci.*, 1996, **31**, 779–784.
44  Y. Li, S. Xie, W. Zhou, L. Ci and Y. Bando, *Chem. Phys. Lett.*, 2002, **356**, 325–330.
45  G. A. Bootsma, W. F. Knippenberg and G. Verspui, *J. Cryst. Growth*, 1971, **11**, 297–309.
46  L. Zhang, H. Zhuang, C. L. Jia and X. Jiang, *CrystEngComm*, 2015, **17**, 7070–7078.
47  V. Reguero, B. Alemán, B. Mas and J. J. Vilatela, *Chem. Mater.*, 2014, **26**, 3550–3557.
48  C. G. Granqvist and R. A. Buhrman, *J. Appl. Phys.*, 1976, **47**, 2200–2219.
49  M. Rana, A. Pendashteh, R. S. Schäufele, J. Gispert and J. J. Vilatela, *Adv. Energy Mater.*, 2022, **12**, 2103469.
50  X. Rodiles, V. Reguero, M. Vila, B. Alemán, L. Arévalo, F. Fresno, V. A. de la P. O'Shea and J. J. Vilatela, *Sci. Rep.*, 2019, **9**, 9239.
51  J. A. Rajesh and A. Pandurangan, *J. Nanosci. Nanotechnol.*, 2014, **14**, 2741–2751.
52  H. Young Kim, J. Park and H. Yang, *Chem. Commun.*, 2003, 256–257.
53  Y. Shi, *Acc. Chem. Res.*, 2015, **48**, 163–173.
54  X. Liu, J. Zhang, A. Vazquez, D. Wang and S. Li, *J. Phys. Chem. A*, 2019, **123**, 10520–10528.
55  H. T. Chiu and J. S. Hsu, *Thin Solid Films*, 1994, **252**, 13–18.
56  W. J. Bullock, R. Walsh and K. D. King, *J. Phys. Chem.*, 1994, **98**, 2595–2601.
57  M. Vazquez-Pufleau, Y. Wang, P. Biswas and E. Thimsen, *J. Chem. Phys.*, 2020, **152**, 24304.
58  R. S. Schäufele, M. Vazquez-Pufleau, A. Pendashteh and J. J. Vilatela, *Nanoscale*, 2022, **14**, 55–64.
59  M. Vazquez-Pufleau and M. Yamane, *Chem. Eng. Sci.*, 2020, **211**, 115230.
60  F. Slootman and J. C. Parent, *J. Aerosol Sci.*, 1994, **25**, 15–21.
61  O. Olsvik, O. A. Rokstad and A. Holmen, *Chem. Eng. Technol.*, 1995, **18**, 349–358.
62  M. Cui, H. Lu, H. Jiang, Z. Cao and X. Meng, *Sci. Rep.*, 2017, **7**, 41990.
63  M. Ghasemi, Z. Zanolli, M. Stankovski and J. Johansson, *Nanoscale*, 2015, **7**, 17387–17396.
64  G. Wilde, *Adv. Eng. Mater.*, 2021, **23**, 2001387.
65  G. Meng, L. Zhang, Y. Qin, F. Phillipp, S. Qiao, H. Guo and S. Zhang, *Chinese Phys. Lett.*, 1998, **15**, 689–691.
66  B. Park, Y. Ryu and K. Yong, *Surf. Rev. Lett.*, 2004, **11**, 373–378.
67  X. Zhang, Y. Chen, Z. Xie and W. Yang, *J. Phys. Chem. C*, 2010, **114**, 8251–8255.
68  X. T. Zhou, N. Wang, H. L. Lai, H. Y. Peng, I. Bello, N. B. Wong, C. S. Lee and S. T. Lee, *Appl. Phys. Lett.*, 1999, **74**, 3942–3944.

# Supplementary information

## *Ultrafast synthesis of SiC nanowire webs by floating catalysts rationalised through in-situ measurements and thermodynamic calculations*

Isabel Gómez-Palos,[†a,b] Miguel Vazquez-Pufleau,[†,a] Jorge Valilla,[a,b] Álvaro Ridruejo,[b] Damien Tourret,[a,*] and Juan J. Vilatela[a,*]

[a]IMDEA Materials, Madrid, 28906, Spain. [b]Department of Materials Science, Universidad Politécnica de Madrid, E.T.S. de Ingenieros de Caminos, 28040 Madrid, Spain. [c]Universidad Carlos III de Madrid, 28911 Leganes, Spain. † These authors contributed equally to this work.

*damien.tourret@imdea.org, juanjose.vilatela@imdea.org.

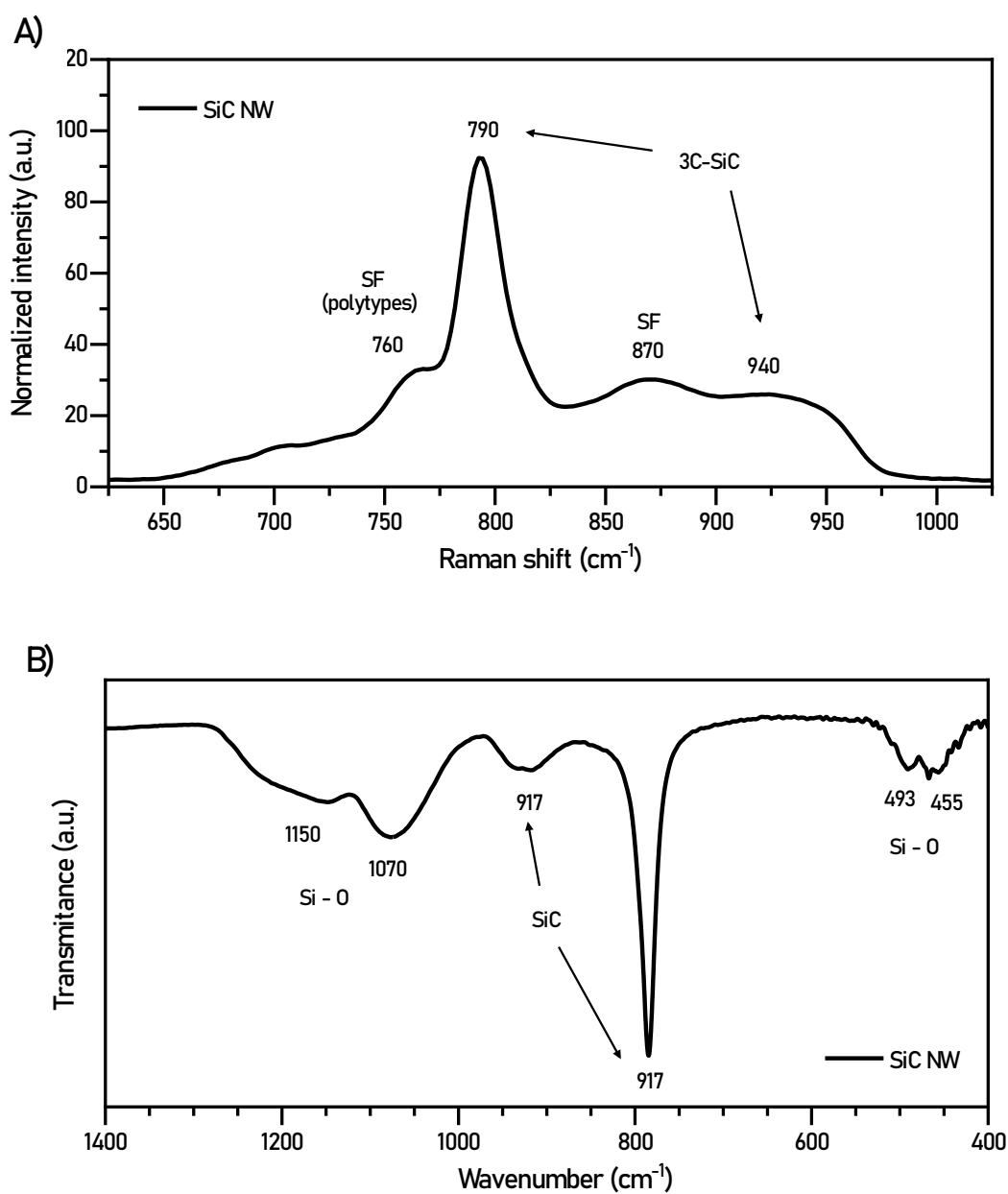

**Figure S1** A) Raman shift of SiC nanowires presenting anomalies caused by nanosegments of different polytypes produced during growth because of stacking faults (SF). B) FTIR spectra spectra of freestanding SiC nanowires web.

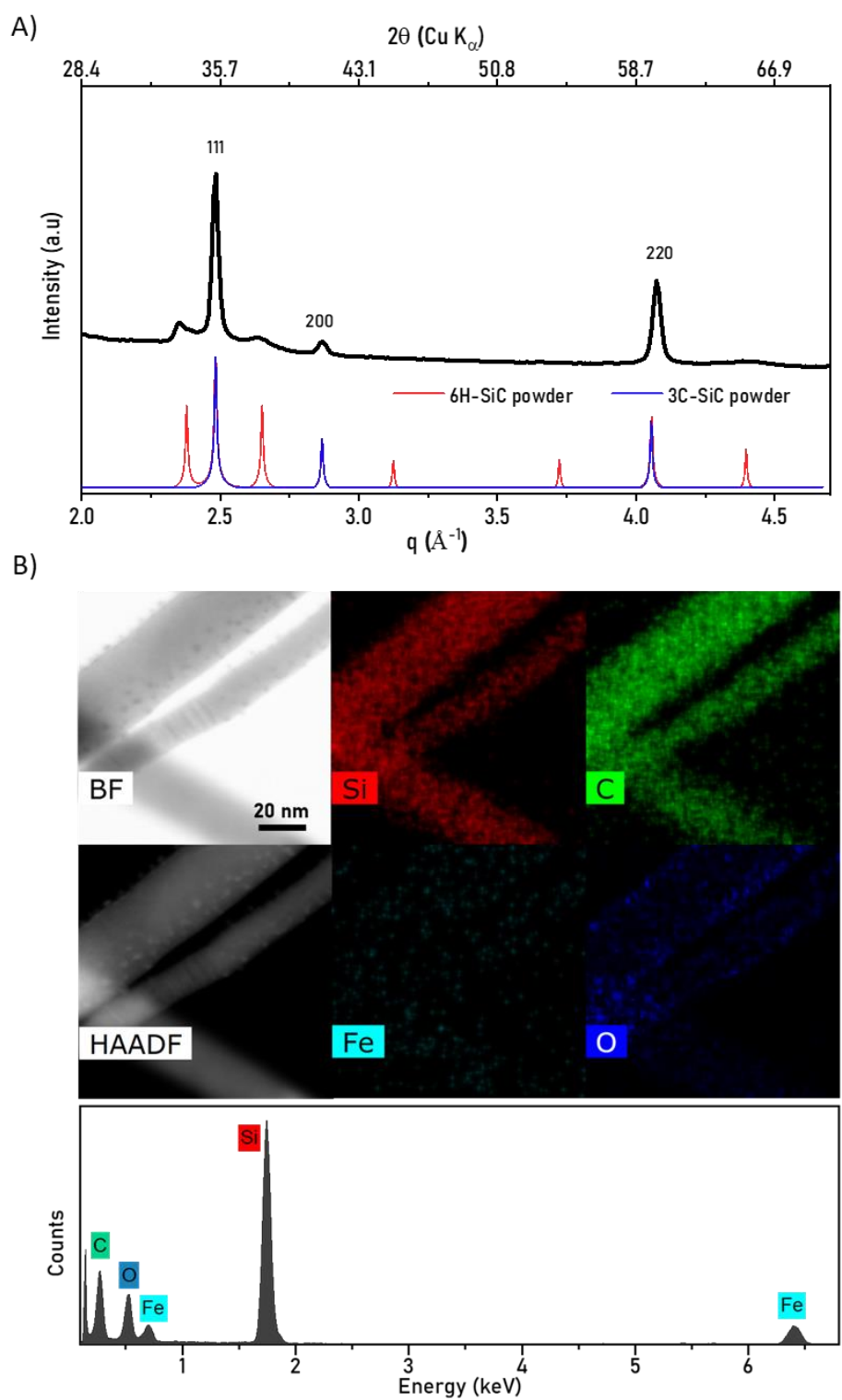

**Figure S2A)** Synchrotron X-Ray diffraction pattern. Sample (black) compared with the powder spectrum of 6H-SiC (red) and 3C-SiC (blue) phase. **B)** EDX map of SiC nanowires confirming that they are composed of Si and C, with only a thin O shell layer from the native oxide formed from exposure to air. (Fe signal in the spectrum comes from catalyst around the sample)

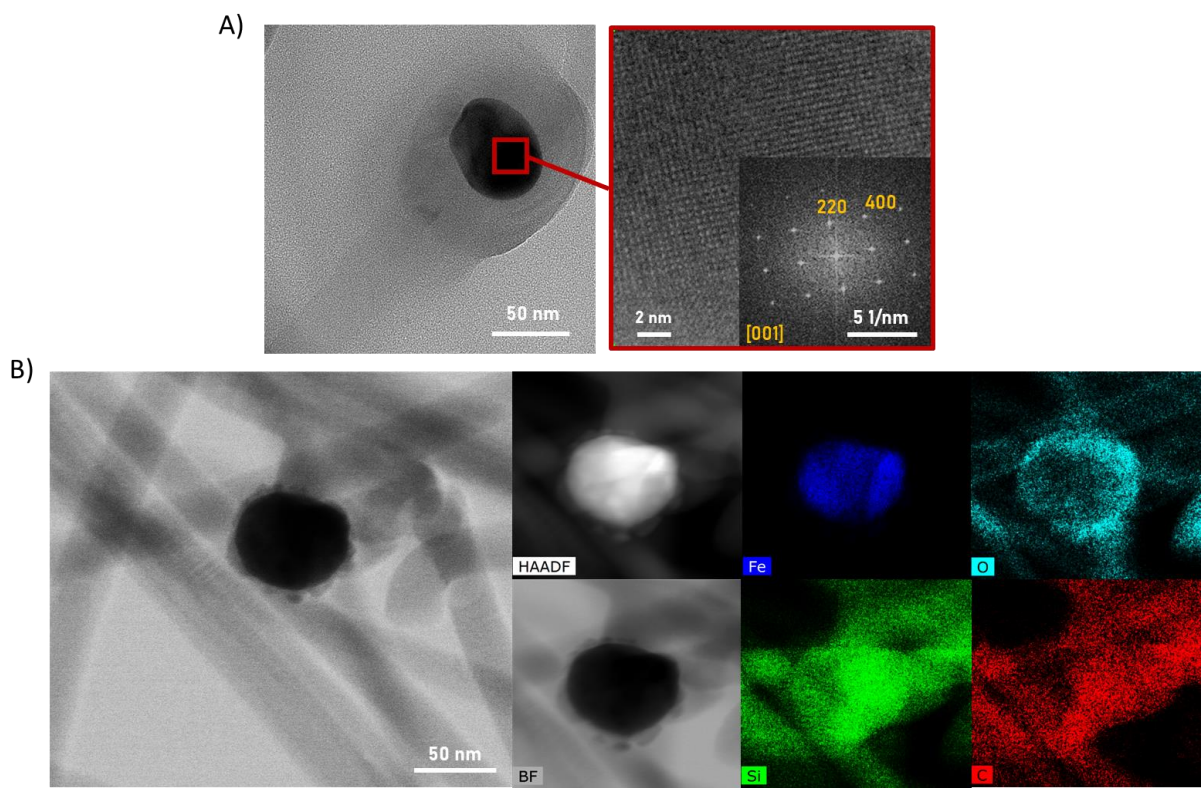

**Figure S3.** HRTEM images of two different catalyst particles A) FFT showing a lattice spacing of of 0.198 nm (220) and 0.144 nm (400), corresponding to the intermetallic Fe$_3$Si phase[1]  B) Catalyst composition map showing Fe and Si as the conforming elements and a surrounding oxide layer.

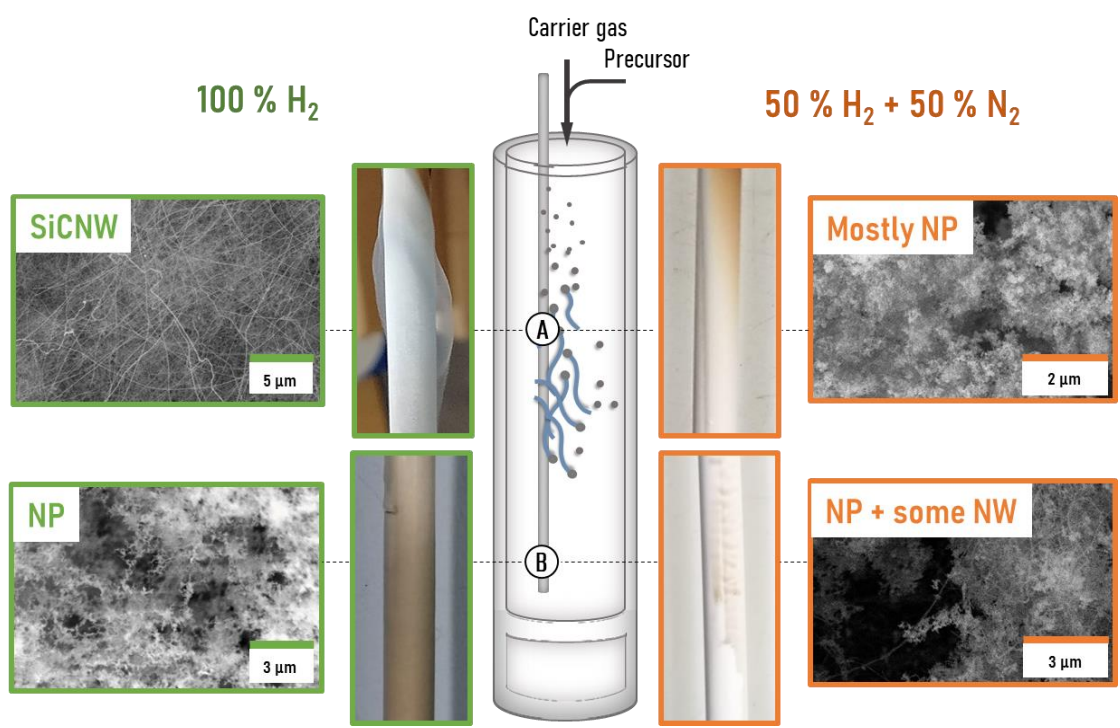

**Figure S4**   Samples recollected with the mullite rod at 30 cm (position A) and 60 cm (position B) from experiments performed at same conditions but with different concentration of N$_2$ in the carrier gas, both at total flow of 3 lpm. For pictures highlited in green, pure H$_2$ was used; while for pictures highlighted in organge, the carrier gas was a mixture 50%/50% of H$_2$ and N$_2$.

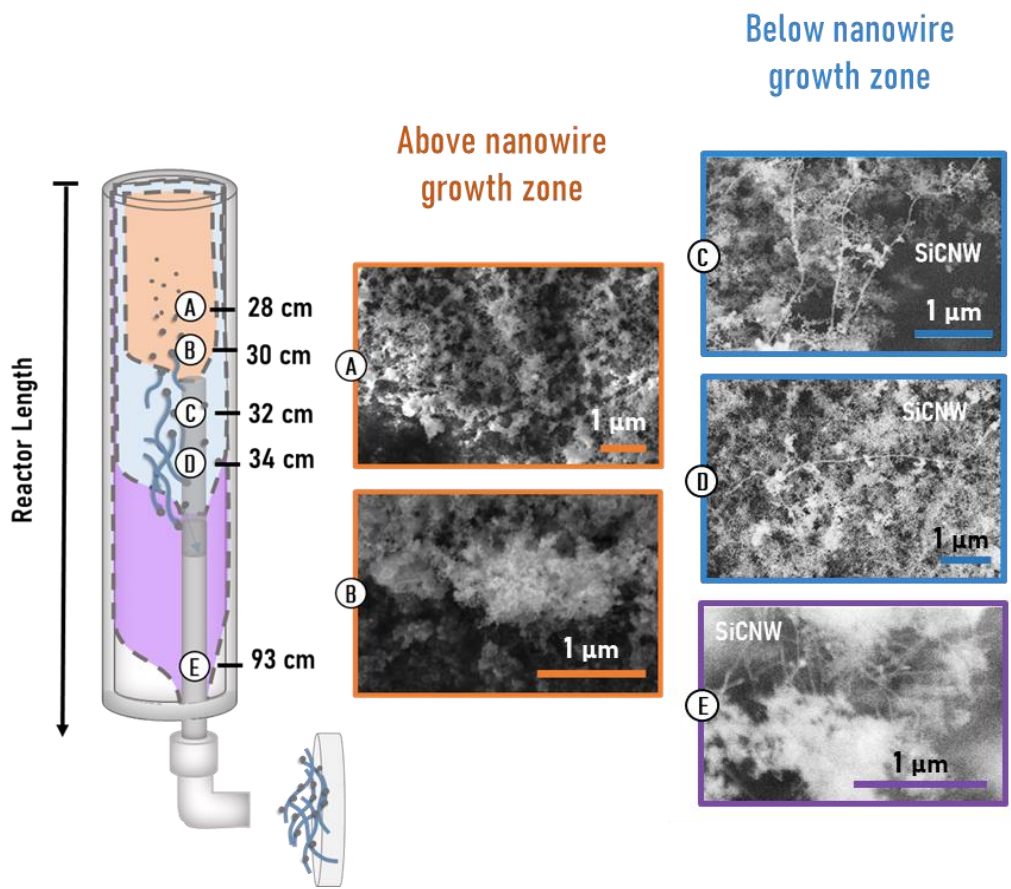

**Figure S5** Samples extracted with vacuum collection system at different positions of the reactor, using catalyst produced by ferrocene decomposition. A-B) Samples taken above the nanowire formation zone do not present nanowires, only nanoparticles produced from pyrolysis. C-E) Samples collected in the nanowire zone or bellow present nanowires along with other by-products.

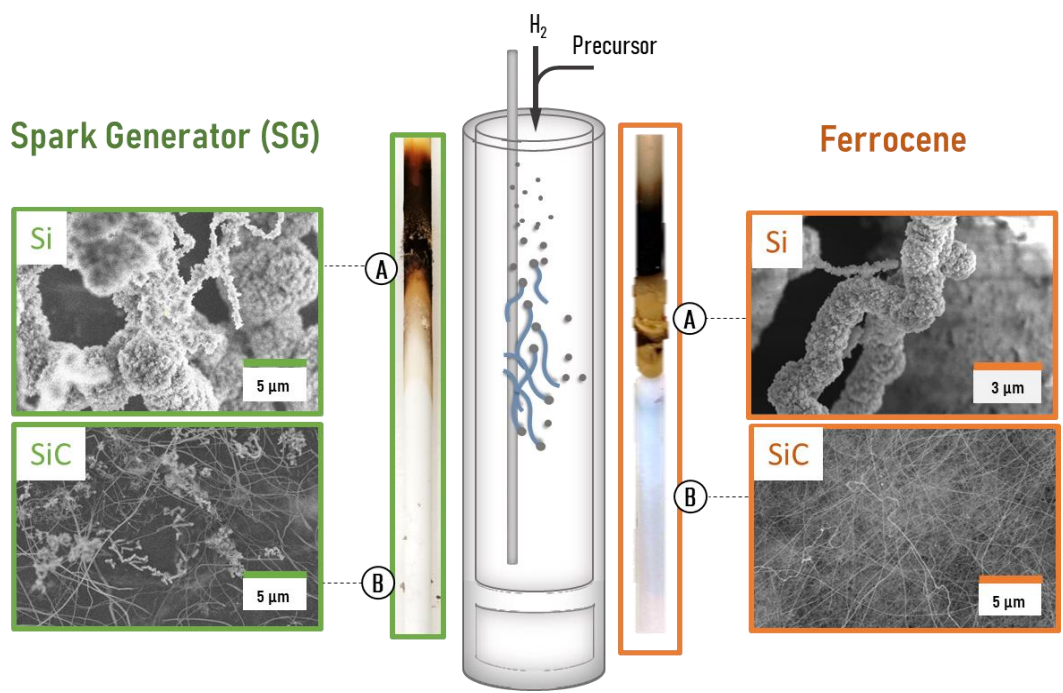

**Figure S6** Samples collected from experiments performed at same conditions but different catalyst source, namely, ferrocene and SG. The samples collected with both sources are fairly similar, A) Si microstructures B) SiC NW.

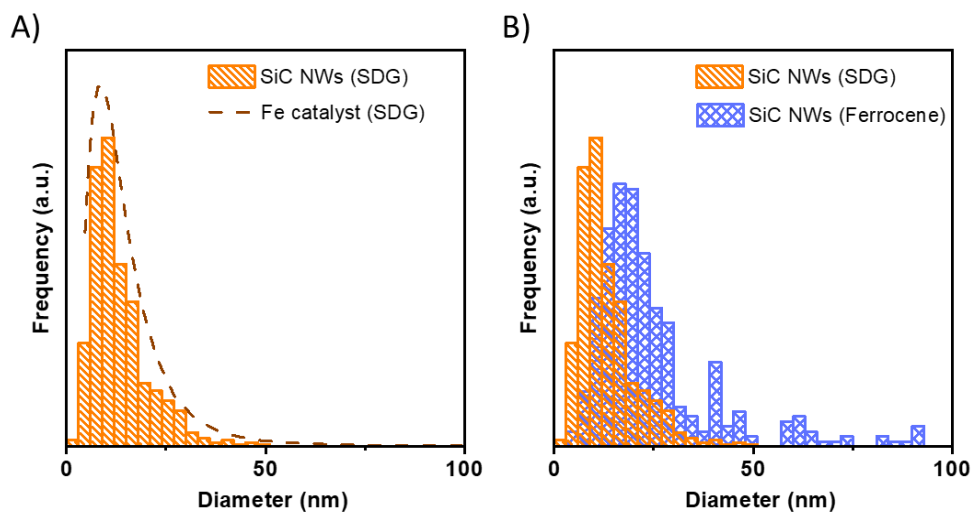

**Figure S7.** A) Comparison of the diameter distribution of the Fe catalyst produced with SDG (dashed line) obtained with in-situ SMPS analysis at the reaction zone, and the diameter distribution of the nanowires grown with this catalyst source (histogram) obtained from image analysis .The SiC NW diameter matches quite well with the floating SDG Fe catalyst. B) Diameter distribution of NWs grown with different catalyst sources, spark discharge generator (orange) and Ferrocene (blue).

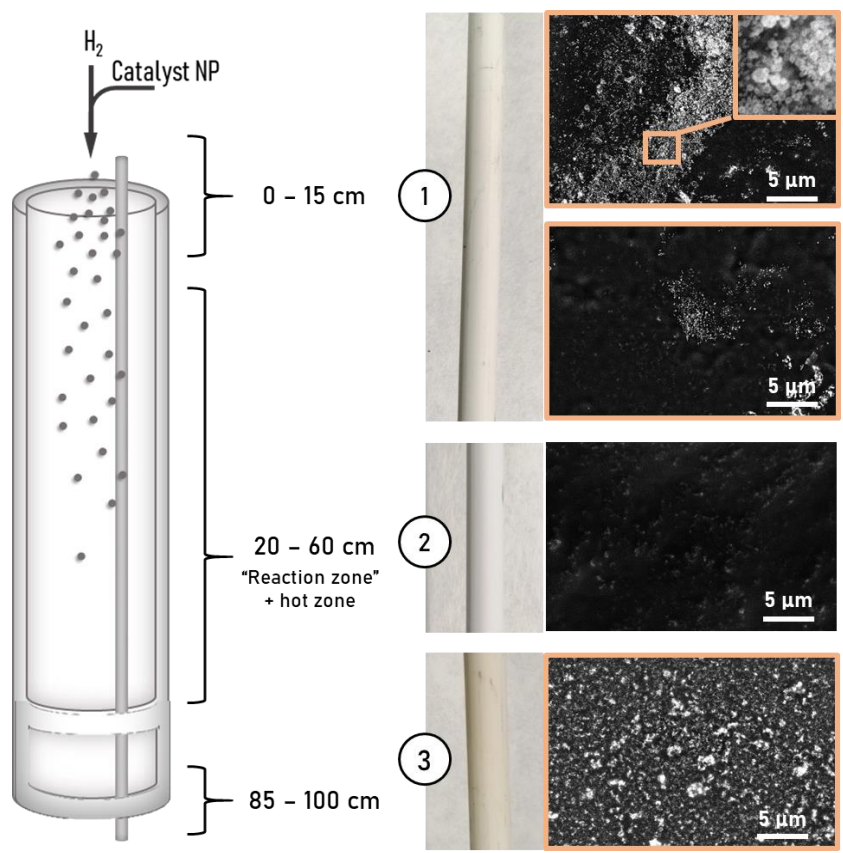

**Figure S8.** Experiment to probe the possible deposition on the ceramic rod of Fe particles generated with SDG. Particles were only detected at positions 1 and 3, corresponding to temperatures below 600 °C. No evidence of attached particles were found on the section of the ceramic rod at a tempearure corresponding to the "Reaction zone"

**Table S1** VLS Growth rate comparison between substrate assisted CVD and this work

| Synthesis method | Temp. and carrier gas | Precursors | Catalyst | Special observations | Aspect ratio | Growth time (min) | Growth rate (nm/s) | Reference |
|---|---|---|---|---|---|---|---|---|
| CVD | 1200 °C, $H_2$ | $SiCl_2$ (g) + Active carbon (s) | Fe | Grown in activated carbon pores | 750 | 90 | 1.4 | Meng 1998 [2] |
| Hot filament CVD | 1000 °C, $H_2$ | Si (s) + graphite (s) | Ni, Cr, Fe impurities | Low pressure | 50 | 120 | 0.1 | Zhou 1999 [3] |
| Thermal heating (SLS) | 1100 °C, Ar | Si substrate (s) + graphite (s) | Ni | $WO_3$ to reduce graphite | 2143 | 180 | 6.9 | Park 2004 [4] |
| CVD | 1300 °C, Ar | l-PS (l) + Active carbon (s) | Fe | Polymer precursor | 66667 | 180 | 925.9 | Li 2009 [5] |
| CVD | 1150 °C, $H_2$ | HMDS (l) | Ni |  | 500 | 30 | 13.9 | Panda 2010 [6] |
| CVD | 1500 °C | Si (s) + graphite (s) | Fe |  | 333 | 360 | 2.3 | Wu 2012 [7] |
| CVD | 1250 °C, $H_2$ | SH4 (g) + $CH_4$ (g) | Fe |  | 1000 | 5 | 166.7 | Attolini 2014 [8] |
| Hot filament CVD | 1400 °C, Ar | Si (s) + Active carbon (s) | $LaNi_5$ | Catalyst enhancer | 375 | 120 | 1.0 | Rajesh 2014 [9] |
| CVD | 1100 °C, $H_2$ | Si substrate (s) + $CH_4$ (g) | Fe, Ga, GaN | Catalyst enhancer | 12500 | 45 | 185.2 | Kim 2002 [10] |
| Thermal heating | 1550 °C, Ar | Polysilazane + graphite substrate (s) | Al | low pressure | 300 | 30 | 16.7 | Zhang 2010 [11] |
| FCCVD | 1200 °C, $H_2$ | HMDS (l) | Fe | Floating catalyst- continuous synthesis | 1871 | 0.15 | 3646 | This work |

**Table S2.** Calculated composition of points (a) through (e) in Fig. 6b, in atomic fraction. Point (c) is calculated from thermodynamic equilibrium of Liquid with FCC phase at composition (b), while compositions (b) and (d) are calculated by addition of C and Si from points (a) and (c), respectively, with a C:Si ratio of 1:3 corresponding to HMDS.

|   | Phase   | X(Fe) | X(Si)  | X(C)   |
|---|---------|-------|--------|--------|
| a | FCC Fe  | 1.0   | 0.0    | 0.0    |
| b | FCC Fe  | 0.906 | 0.0235 | 0.0705 |
| c | Liquid  | 0.832 | 0.0202 | 0.1475 |
| d | Liquid  | 0.814 | 0.0248 | 0.1615 |
| e | Liquid  | 0.693 | 0.273  | 0.034  |

**Length Measurement:**

Determining the length of high aspect ratio nanomaterials is established as a long-standing challenge. Despite extensive research on CNTs produced by FCCVD, the only reported estimates of length are a calculation based on the number of ends found in a sample,[12] and indirect measurements using polarised IR spectroscopy.[13] The high aspect ratio of the nanowires in this work makes TEM unsuitable for length measurements. Observation under TEM requires the material to be very thin, essentially as individualised NWs. This implies that they need to be fully dispersed in solvents and then deposited on TEM grids, without shortening them. Previous work on CNTs, SiNWs and now SiCNWs has proven this method unsuccessful. Instead, taking advantage of the relatively large diameter of SiCNWs, we directly image them by high resolution SEM. In this pristine sate though, the NWs are highly entangled, requiring extensive observation to identify NWs with their whole length exposed.

To measure the mean length we used a web collected with the ceramic rod which was then introduced in isopropanol, subjected to mild sonication for 3 minutes. After sonication, a droplet of the material was deposited on top of a flat substrate. Multiple high-resolution images were taken and joined together to create a high quality maps of the nanowire networks, as shown in the example in Figure 3c. Next, the resulting image was visually inspected to identify nanowires whose entire length could be resolved, typically corresponding to those on top of the network and/or near its edges. Their length was determined using Image J. Fig S9 shows an example. This process was applied to five different samples, and a total of 170 nanowires.

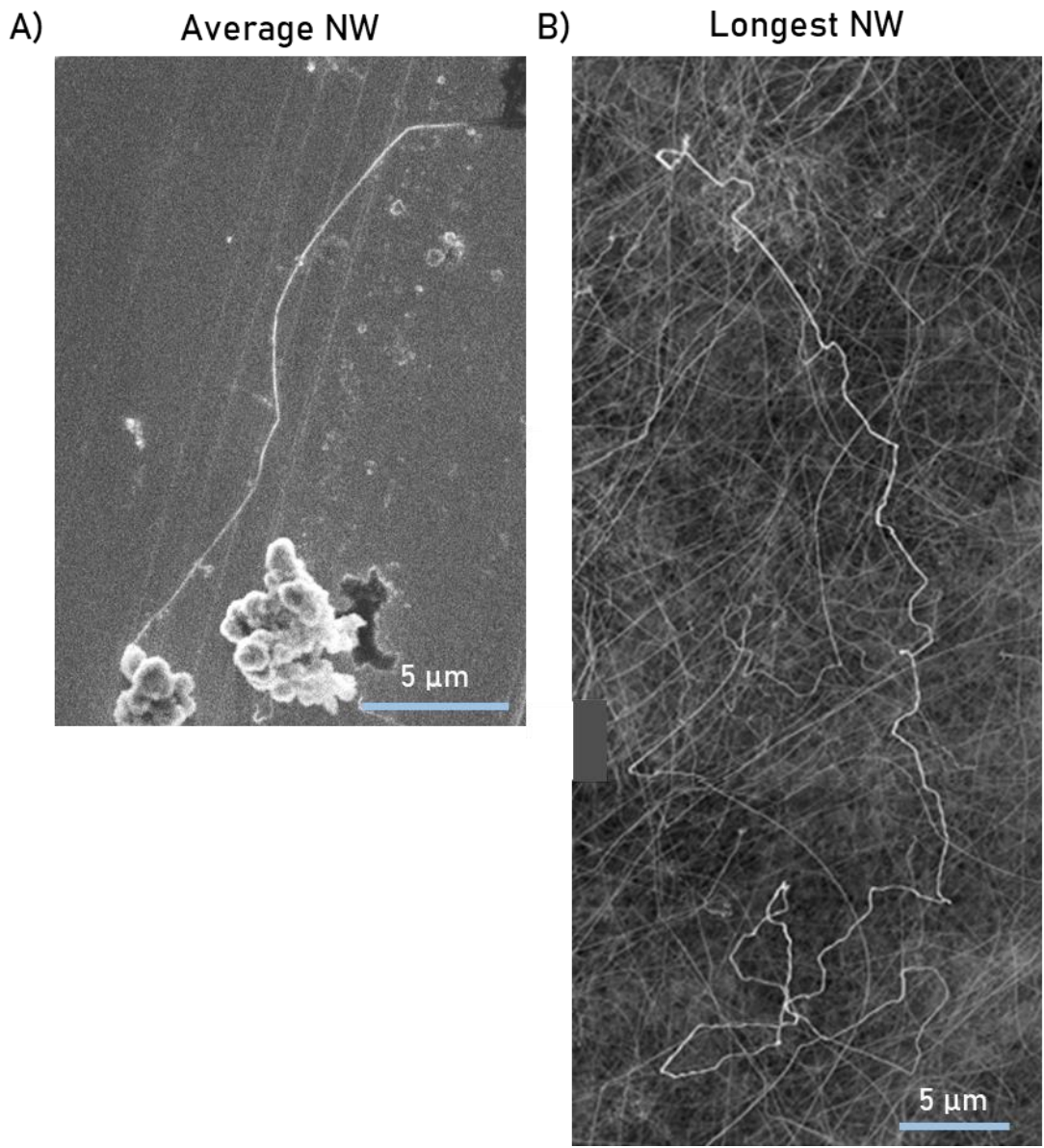

**Figure S9**. Example of a nanowire with average length and diameter (a) and a SEM micrograph map of the longest nanowire found in sonicated samples synthetised at standard conditions (L = 123 µm, D = 55 nm).

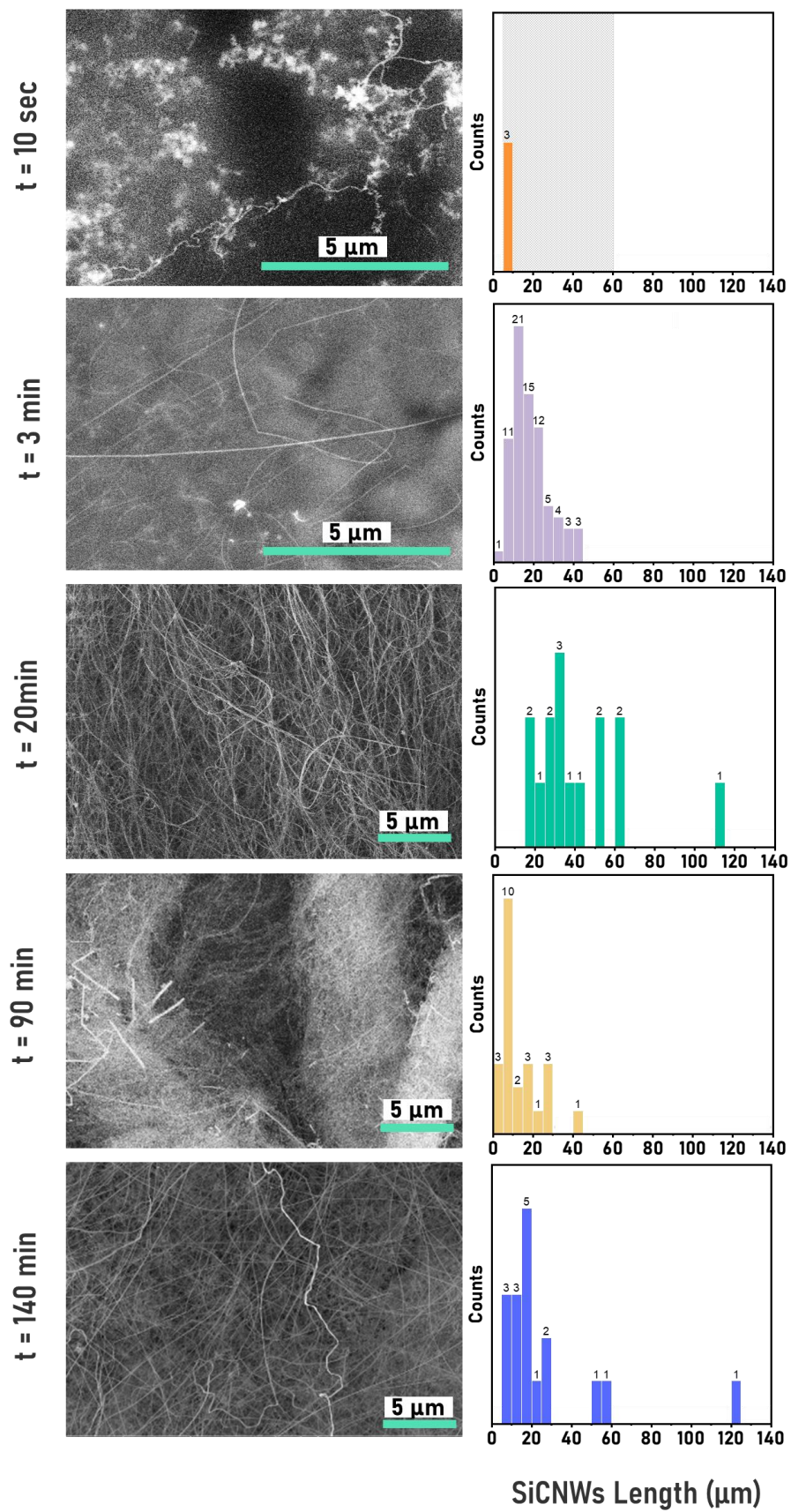

**Figure S10.** Length of NW collected by the ceramic rod at different experiment times. Standard synthesis conditions were used in all experiments shown.

**By-product analysis:**

Figure S11 shows examples of high resolution TEM images of the principal solid by-product of the reactor, their FFT confirms the composition as c-Si and amorphous Si-C nanoparticles with some SiC crystalline phases. Examples of samples collected with ceramic rods at different experimental conditions, and the Raman spectra of their different solid products are shown in Figure S12.

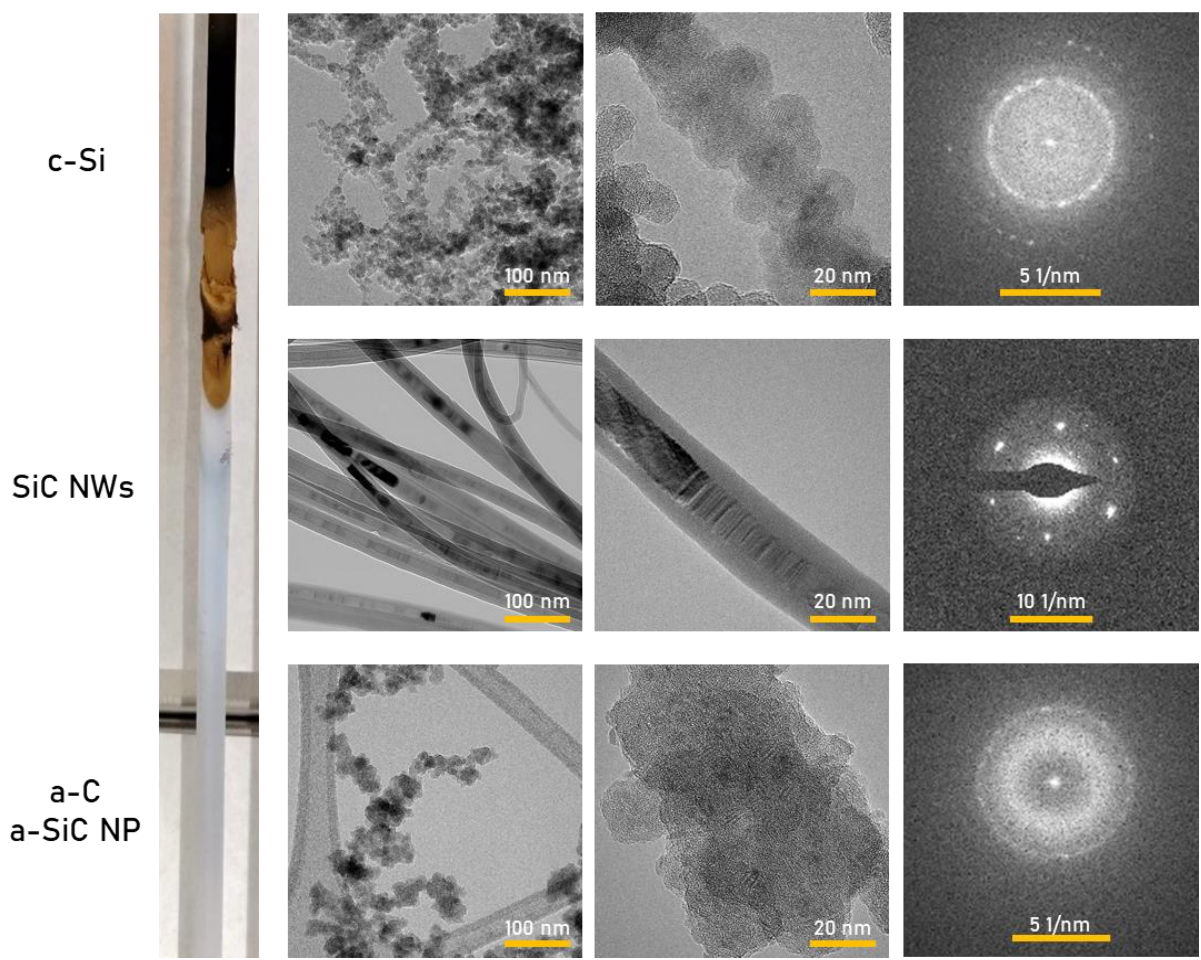

**Figure S11.** TEM micrographs of the different solids produced at different reaction temperatures and collected on a ceramic rod instered in the reactor tube: Si nanoparticle aggregates, SiC nanowires and SiC/C soot.

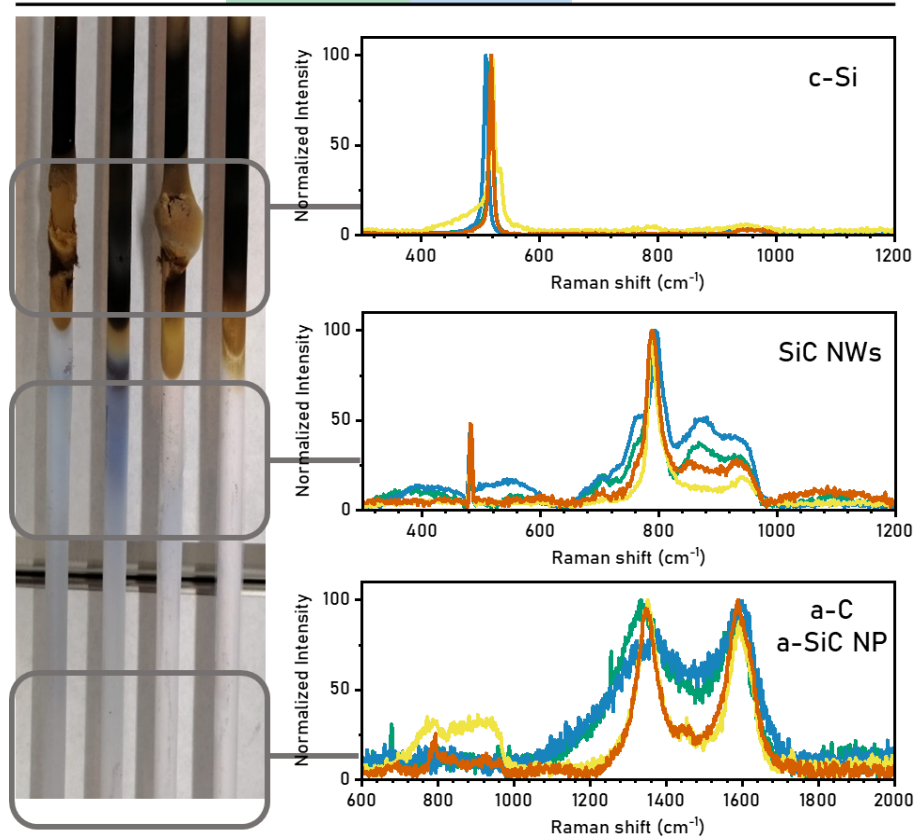

**Figure S12.** Raman spectra of the different solids produced at different reaction temperatures and collected on a ceramic rod instered in the reactor tube. The composition of the solid products is the same for the range of synthesis conditions studied (Ferroce or SG as catalyst source, 1300 C or 1250 °C, and carrier gas with 10% of nitrogen), although their concentrations may differ.

**Sample Collection systems**

The ceramic rod and vacuum filter used in this work are expected to have very different collection efficiencies for the different solid particles formed in the reaction. Nanowires, particularly those of high aspect ratio can aggregate, entangle in the gas phase and attach to the surface of the ceramic rod and/or the reactor tube walls. Soot particles are less prone to aggregation and instead more likely to be collected by the vacuum system. Hence the observation that the material collected on the ceramic cold finger has a higher fraction of NWs than that collected at the cold end of the reactor with the vacuum system. Save for recent work on the formation of carbon nanotube aerogels,[15,16] the dynamics of high aspect ratio nanoparticles floating in a gas stream is a largely unresolved problem.

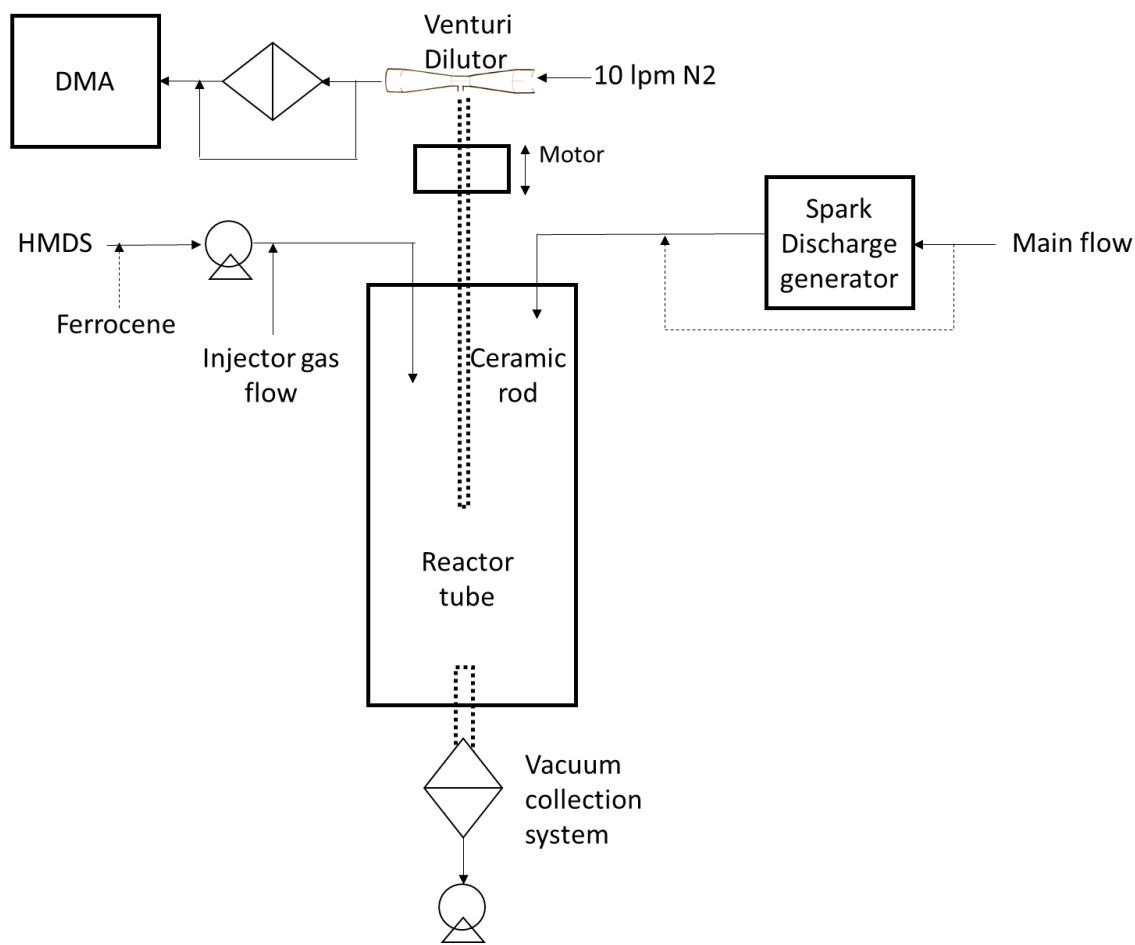

**Figure S13.** Reactor configuration. Pointed lines represent collection systems that are not used at the same time. Dash lines represent the configuration used when ferrocene is used instead of spark generation to produce the Fe catalyst.

References

1    L. Zhang, H. Zhuang, C. L. Jia and X. Jiang, *CrystEngComm*, 2015, **17**, 7070–7078.

2    G. Meng, L. Zhang, Y. Qin, F. Phillipp, S. Qiao, H. Guo and S. Zhang, *Chinese Phys. Lett.*, 1998, **15**, 689–691.

3    X. T. Zhou, N. Wang, H. L. Lai, H. Y. Peng, I. Bello, N. B. Wong, C. S. Lee and S. T. Lee, *Appl. Phys. Lett.*, 1999, **74**, 3942–3944.

4    B. Park, Y. Ryu and K. Yong, *Surf. Rev. Lett.*, 2004, **11**, 373–378.

5   G. Li, X. Li, Z. Chen, J. Wang, H. Wang and R. Che, *J. Phys. Chem. C*, 2009, **113**, 17655–17660.

6   S. K. Panda, J. Sengupta and C. Jacob, *J. Nanosci. Nanotechnol.*, 2010, **10**, 3046–3052.

7   R. Wu, K. Zhou, J. Wei, Y. Huang, F. Su, J. Chen and L. Wang, *J. Phys. Chem. C*, 2012, **116**, 12940–12945.

8   G. Attolini, F. Rossi, M. Negri, S. C. Dhanabalan, M. Bosi, F. Boschi, P. Lagonegro, P. Lupo and G. Salviati, *Mater. Lett.*, 2014, **124**, 169–172.

9   J. A. Rajesh and A. Pandurangan, *J. Nanosci. Nanotechnol.*, 2014, **14**, 2741–2751.

10  H. Young Kim, J. Park and H. Yang, *Chem. Commun.*, 2003, 256–257.

11  X. Zhang, Y. Chen, Z. Xie and W. Yang, *J. Phys. Chem. C*, 2010, **114**, 8251–8255.

12  M. Motta, A. Moisala, I. A. Kinloch and A. H. Windle, *Adv. Mater.*, , DOI:10.1002/adma.200700516.

13  T. Morimoto, S.-K. Joung, T. Saito, D. N. Futaba, K. Hata and T. Okazaki, *ACS Nano*, 2014, **8**, 9897–9904.

14  H. Liu and P. H. Zwart, *J. Struct. Biol.*, 2012, 180, 226–234.

15  N. Kateris, P. Kloza, R. Qiao, J. A. Elliott and A. M. Boies, *J. Phys. Chem. C*, 2020, **124**, 8359–8370.

16  A. M. Boies, C. Hoecker, A. Bhalerao, N. Kateris, J. de La Verpilliere, B. Graves and F. Smail, *Small*, 2019, **15**, 1900520.